\documentclass[aps,preprint,amsmath,amssymb]{revtex4}
\usepackage{graphicx}
\usepackage{epstopdf}

\begin{document} %

\title{Search for the RS model with a small curvature through photon-induced process at the LHC}

\author{S.C. \.{I}nan}
\email[]{sceminan@cumhuriyet.edu.tr}
\affiliation{Department of Physics, Cumhuriyet University,
58140, Sivas, Turkey}
\author{A.V. Kisselev}
\email[]{alexandre.kisselev@ihep.ru} \affiliation{A.A. Logunov
Institute for High Energy Physics, NRC ``Kurchatov Institute'',
142281 Protvino, Russian Federation}


\begin{abstract}
In this paper, potential of the LHC to explore the phenomenology of
the Randall-Sundrum-like scenario with the small curvature for the
process $pp \to p \gamma \gamma p \to p \mu^- \mu^+p $ through the
subprocess $\gamma \gamma \to \mu^- \mu^+$ is examined for two
forward detector acceptances, $0.0015 < \xi< 0.5$ and $0.1< \xi
<0.5$. This process is known to be one of the most clean channels.
The sensitivity bounds on the anomalous model parameters have been
found at the $95\%$ confidence level for various LHC integrated
luminosity values.
\end{abstract}

\maketitle


\section{Introduction} %

The Standard Model (SM) gives very satisfactory description of high
energy physics at an energy scale of electroweak interactions. It is
in perfect agreement with current experimental results. However,
there are many open equations at the SM. One of this is the
hierarchy problem. The extra dimensional models in high energy
physics provide possible candidates for this problem.  In this
respect, these models have drawn attention over the recent years.
There are many phenomenology papers of new physics models at the
LHC. These searches generally include the usual proton-proton
inelastic processes in which the interacted protons dissociate into
jets. Due to these jets, such interactions give a very crowded
environment. These formed jets create some uncertainties and make it
difficult to detect the signals from the new physics which is beyond
the SM. However, exclusive production $pp \to pXp$ provides a very
clean environment. These type of processes have been much less
examined in the literature. Both interacted protons remain intact,
hence they do not dissociate into partons in the exclusive
productions. ATLAS and CMS collaborations prepared a physics program
of forward physics with extra detectors symmetrically located in a
distance from the interaction point. These new detectors are
equipped with charged particle trackers and they provide to tag
intact scattered protons after the collision. Additionally, forward
detectors can detect intact outgoing protons in the interval
$\xi_{min}<\xi<\xi_{max}$ where $\xi$ is the momentum fraction loss
of the intact protons $\xi=(|E|-|E^{\,\,\prime}|)/|E|$ where $E$ and
$E^{\,\,\prime}$ are the energies of the incoming and intact
scattered proton, respectively. These intervals are known as the
acceptance of the forward detectors. If these machines are located
closer to main detectors, a higher $\xi$ can be created. With
applying these new detectors it is possible to obtain high energy
photon-photon process with exclusive two particle final states such
as leptons or photons. The programs about these detectors were
prepared by ATLAS Forward Physics Collaboration (AFP) and CMS-TOTEM
Precision Proton Spectrometer (CT-PPS) \cite{afp, totem}. AFP cover
$0.0015 < \xi <0.15$, $0.015 < \xi <0.15$ detector acceptance
ranges. Similarly, CT-PPS has the acceptance ranges of  $0.0015 <
\xi< 0.5$, $0.1< \xi <0.5$. Two types of measurements are planned by
AFP with high precision: i) Exploratory physics (anomalous couplings
between $\gamma$ and $Z$ or $W$ bosons, exclusive production, etc.),
ii) standard QCD physics (double Pomeron exchange, exclusive
production in the jet channel, single diffraction, $\gamma\gamma$
physics, etc.) \cite{afp1, afp2}. The CT-PPS experiment main
motivations are the investigation of the proton-proton total
cross-section, elastic proton-proton interactions, and all of the
diffractive processes. Roman Pots detector is used in this
experiment. In the forward location, almost all inelastic physical
interactions can be detected by the charged particle detectors. A
large solid angle unable to cover with the support of the CMS
detector. In this way, the detectors can be used for precise studies
\cite{ttm1, ttm2, ttm3}. Due to high energy and high luminosity,
this kind of interactions may cause a number of pile-up events.
However, these backgrounds enable to be rejected by applied
exclusivity conditions, kinematics and timing constraints with use
of forward detectors in conjunction with central detectors.
Moreover, two lepton final states have very small backgrounds in the
presence of pile-up events. Because there are no other charged
particles on the two lepton interaction vertex. Therefore, final
state leptons are highly back-to-back with almost equivalent $p_t$
\cite{albrow, albrow1}.

Photon induced reactions were studied by the CDF collaboration
\cite{cdf1,cdf2}. Obtained results in these experiments are
consistent in theoretical expectations with $p\bar{p} \to
p\ell^{-}\ell^{+}\bar{p}$ through the subprocess $\gamma\gamma \to
\ell^{-}\ell^{+}$. At the LHC, the CMS collaboration have recently
examined to measurement of the photon-induced reactions through the
processes $pp \to p\gamma \gamma p \to p\mu^+ \mu^- p $, $pp \to
p\gamma \gamma p \to p e^- e^+ p $ from the $\sqrt{s}=7$ TeV
\cite{ch1,ch2}. This experiments show that both the number of
candidates and the kinematic distributions are in agreement with the
expectation for exclusive and semi-exclusive $e^-e^+$ production via
$\gamma\gamma \to e^-e^+$ process. Similarly, ATLAS Collaboration
reported a measurement of the exclusive $\gamma\gamma \to l^- l^+$
($l=e,\mu$ ) cross-section in proton-proton collisions at a
center-of-mass energy of $7$ TeV \cite{ath1}. In this measurement,
it is obtained that when proton absorptive effects due to the finite
size of the proton are taken into account, the obtained
cross-sections are found to be consistent with the theoretical
calculations. Another similar measurement was made by the ATLAS
Collaboration at a center-of-mass energy of $8$ TeV \cite{ath2}. In
this experiment, the single-differential cross section was obtained
as a function of $m_{\ell^{-}\ell^{+}}$ ($\ell=e, \mu$). The leptons
channel measurements are combined and a total experimental precision
of better than $1\%$ is achieved at low $m_{\ell \ell}$. In the
literature, other phenomenological papers based on photon-induced
reactions at the LHC for new physics beyond the SM can be found in
\cite{lhc2,lhc4,inanc,inan,bil,bil2,kok,inan2,gru,inanc2,ban,epl,inanc3,bil4,inanc4,hao1,hao2,ins,kok2,fic1,fic2}.

\begin{figure}[htb]
\includegraphics{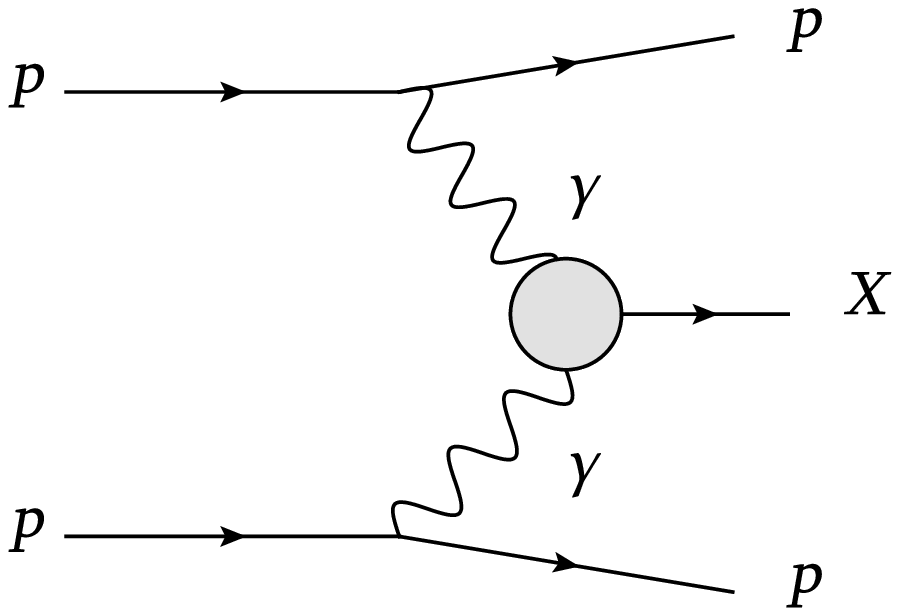}
\caption{Shematic diagram for the reaction $pp\to p \gamma \gamma p \to pXp$.}
\label{sch}
\end{figure}

The forward detectors allow high energy photon-photon interaction as
mention above. The photons which are generated by the high energetic
protons can be considered as an intense photon beam. These
almost-real photons have very low virtuality so that it can be
assumed that they are on-mass-shell. They radiate off the incoming
protons with small angles and low transverse momentum. Intact
protons thus deviate slightly from their trajectory along the beam
path without being detected by central detectors. Intact protons and
$\xi$ are measured by the forward detectors with a very large
pseudorapidity. As a result, the final state $X$ is obtained through
the process $pp \to p\gamma \gamma p \to pXp $ and measured at the
main detector. The schematic diagram for this process is shown in
Fig.\ref{sch}. Since, energy loses protons can be detected by the
forward detectors, the invariant mass of the central system
$W=2E\sqrt{\xi_1 \xi_2}$ can be known.

The photon-induced reactions can be described by the equivalent
photon approximation (EPA) \cite{budnev,baur}. According to this
method, quasi-real photons with low vitalities ($Q^2 = -q^2$)
interact to create final state X through the subprocess
$\gamma\gamma \to X$. In the EPA framework, emitted quasi-real
photons bring a spectrum that is a function of virtuality $Q^2$ and
the photon energy $E_\gamma=\xi E$,

\begin{eqnarray}
\frac{dN_\gamma}{dE_{\gamma}dQ^{2}}=\frac{\alpha}{\pi}\frac{1}{E_{\gamma}Q^{2}}
\left[ (1-\frac{E_{\gamma}}{E})
(1-\frac{Q^{2}_{min}}{Q^{2}})F_{E}+\frac{E^{2}_{\gamma}}{2E^{2}}F_{M}
\right] . \label{phs}
\end{eqnarray}

\noindent Here $m_{p}$ proton mass. In the dipole approximations the
other terms given as follows,

\begin{eqnarray}
Q^{2}_{min}=\frac{m^{2}_{p}E^{2}_{\gamma}}{E(E-E_{\gamma})},
\;\;\;\; F_{E}=\frac{4m^{2}_{p}G^{2}_{E}+Q^{2}G^{2}_{M}}
{4m^{2}_{p}+Q^{2}} \;, \\
G^{2}_{E}=\frac{G^{2}_{M}}{\mu^{2}_{p}}=(1+\frac{Q^{2}}{Q^{2}_{0}})^{-4},
\;\;\; F_{M}=G^{2}_{M}, \;\;\; Q^{2}_{0}=0.71 \mbox{\ GeV}^{2}.
\end{eqnarray}

\noindent where, $\mu_{p}^2=7.78$ is the square of the magnetic
moment of the proton, $F_{E}$ and $F_{M}$ are the relative the
electric and magnetic form factors of the proton, respectively. In
the photon-photon collisions, luminosity spectrum
$\frac{dL^{\gamma\gamma}}{dW}$ can be found with using EPA as
follows,

\begin{eqnarray}
\label{efflum}
\frac{dL^{\gamma\gamma}}{dW}=\int_{Q^{2}_{1,min}}^{Q^{2}_{max}}
{dQ^{2}_{1}}\int_{Q^{2}_{2,min}}^{Q^{2}_{max}}{dQ^{2}_{2}} \int_{y_{
min}}^{y_{max}} {dy \frac{W}{2y} f_{1}(\frac{W^{2}}{4y}, Q^{2}_{1})
f_{2}(y,Q^{2}_{2})} \;,
\end{eqnarray}

\noindent with $y_{min}=\mbox{max}(W^{2}/(4\xi_{max}E), \xi_{min}E),
\;\;\; y_{max}=\xi_{max}E, \;\;\; f=\frac{dN}{dE_{\gamma}dQ^{2}}.$
Here, $Q_{max}^2=2$ GeV$^2$ is taken in the above equation since the
contribution of higher virtualities more than this value is
negligible. The cross section for the $pp \to p \gamma \gamma p \to
p X p $ can be derived by integrating selected subprocess $\gamma
\gamma \to X$  cross section over the photon spectrum,

\begin{eqnarray}
\label{completeprocess}
 d\sigma=\int{\frac{dL^{\gamma\gamma}}{dW}
\,d\hat {{\sigma}}_{\gamma\gamma \to X}(W)\,dW}.
\end{eqnarray}

\begin{figure}[htb]
\begin{center}
\includegraphics[scale=0.60]{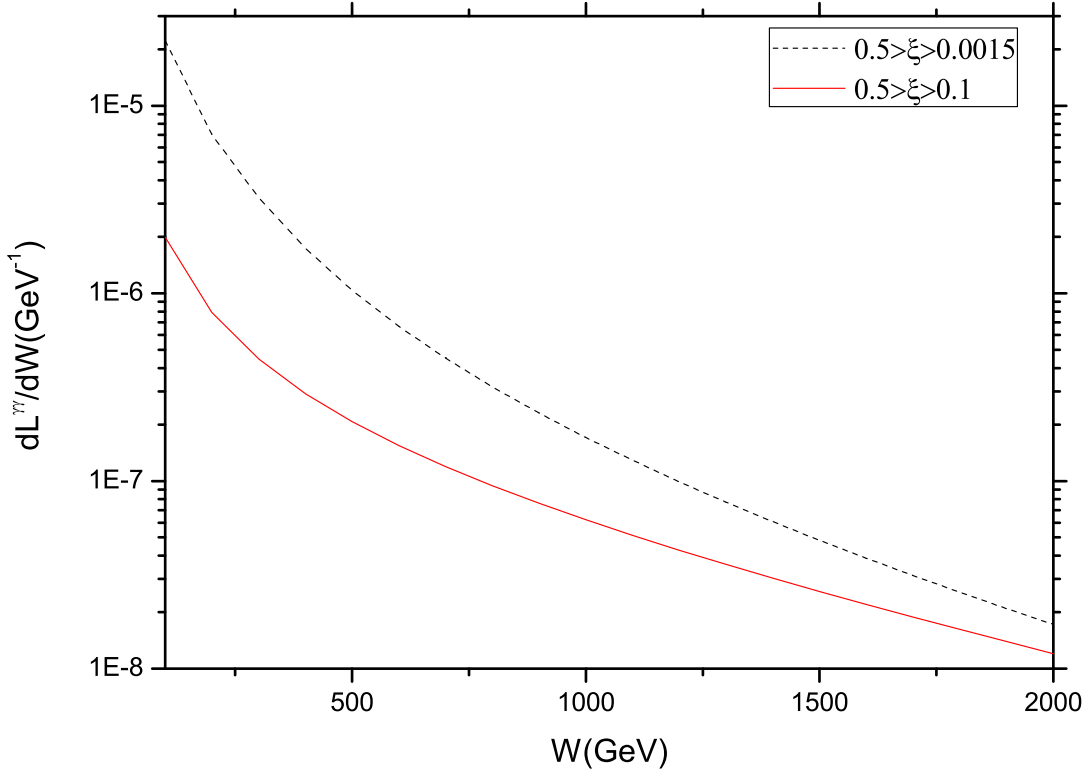}
\caption{Effective $\gamma\gamma$ luminosity as a function of the invariant
mass of the two photon system. Figure shows the effective luminosity
for two forward detector acceptances: $0.0015 < \xi < 0.5$ and $0.1
< \xi < 0.5$.}
\label{fig:lum}
\end{center}
\end{figure}

In this paper, we have investigated the RS-like scenario with the
small curvature (the details is given in the next section) for the
process $pp \to p \gamma \gamma p \to p \mu^- \mu^+p $ through the
subprocess $\gamma \gamma \to \mu^- \mu^+$. Because of the main
contribution comes from the high energy region, we have made this
calculation for two acceptance ranges $0.0015<\xi<0.5$ and
$0.1<\xi<0.5$. Fig.~\ref{fig:lum} shows the behavior of the
$\gamma\gamma$ luminosity spectrum as a function of the $W$ for
these two forward acceptance ranges.

\section{RSSC model of warped extra dimension with small curvature} %

The Randall-Sundrum model with two branes (RS1 model
\cite{Randall:1999}) was proposed as an alternative to the scenario
with large flat extra dimensions (ADD model
\cite{Arkani-Hamed:1998,Hamed:1998,Hamed2:1998}). The RS1 model is described by the
following background warped metric

\begin{equation}\label{RS_metric}
\quad ds^2 = e^{-2 \sigma (y)} \, \eta_{\mu \nu} \, dx^{\mu} \,
dx^{\nu} - dy^2 \;,
\end{equation}

\noindent where $\eta_{\mu\nu}$ is the Minkowski tensor with the signature
$(+,-,-,-)$, and $y$ is an extra coordinate. The periodicity
condition $y=y + 2\pi r_c$ is imposed, and the points $(x_\mu,y)$
and $(x_\mu,-y)$ are identified. Thus, we have a model of gravity in
a slice of the AdS$_5$ space-time compactified to the orbifold
$S^1\!/Z_2$. The orbifold has two fixed points, $y=0$ and $y=\pi
r_c$. It is assumed that there are two branes located at these
points (called Planck and TeV brane, respectively). All the SM
fields are confined to the TeV brane.

The classical action of the RS1 model is given by
\cite{Randall:1999}

\begin{align}\label{action}
S &= \int \!\! d^4x \!\! \int_{-\pi r_c}^{\pi r_c} \!\! dy \,
\sqrt{G} \, (2 \bar{M}_5^3 \mathcal{R}
- \Lambda) \nonumber \\
&+ \int \!\! d^4x \sqrt{|g^{(1)}|} \, (\mathcal{L}_1 - \Lambda_1) +
\int \!\! d^4x \sqrt{|g^{(2)}|} \, (\mathcal{L}_2 - \Lambda_2) \;,
\end{align}

\noindent where $G_{MN}(x,y)$ is the 5-dimensional metric, with $M,N =
0,1,2,3,4$, $\mu = 0,1,2,3$. The quantities
\begin{equation}
g^{(1)}_{\mu\nu}(x) = G_{\mu\nu}(x, y=0) \;, \quad
g^{(2)}_{\mu\nu}(x) = G_{\mu\nu}(x, y=\pi r_c)
\end{equation}

\noindent are induced metrics on the branes, $\mathcal{L}_1$ and
$\mathcal{L}_2$ are brane Lagrangians, $G = \det(G_{MN})$, $g^{(i)}
= \det(g^{(i)}_{\mu\nu})$. $\bar{M}_5$ is a reduced 5-dimensional
Planck scale, $M_5/(2\pi)^{1/3}$, where $M_5$ is a fundamental
gravity scale. The quantity $\Lambda$ is a 5-dimensional
cosmological constant, $\Lambda_{1,2}$ are brane tensions.

The function $\sigma(y)$ in \eqref{RS_metric} was obtained in
\cite{Randall:1999} to be

\begin{equation}\label{sigma_RS}
\sigma_{\mathrm{RS}}(y) = \kappa |y| \;,
\end{equation}

\noindent where $\kappa$ is a parameter with a dimension of mass. It defines
the curvature of the 5-dimensional space-time. Recently, a
generalization of the warp factor $\sigma(y)$ war derived in
\cite{Kisselev:2016}

\begin{equation}\label{sigma}
\sigma(y) = \frac{\kappa r_c}{2} \left[ \left| \mathrm{Arccos}
\left(\cos \frac{y}{r_c} \right) \right| - \left| \pi -
\mathrm{Arccos} \left(\cos \frac{y}{r_c} \right)\right| \right] +
\frac{\pi \,|\kappa| r_c }{2} - C \;,
\end{equation}

\noindent where $C$ is $y$-independent quantity, with the fine
tuning relations

\begin{equation}\label{fine_tuning}
\Lambda = -24 \bar{M}_5^3\kappa^2 \;, \quad \Lambda_1 = -
\,\Lambda_2 = 24 \bar{M}_5^3 \kappa \;.
\end{equation}

\noindent Here $\mathrm{Arccos(z)}$ is a principal value of the multivalued
inverse trigonometric function $\arccos(z)$. This generalized
solution (i) obeys the orbifold symmetry $y \rightarrow - y$; (ii)
makes the jumps of $\sigma'(y)$ on both branes; (iii) has the
explicit symmetry with respect to the branes \cite{Kisselev:2016}.

By taking $C=0$ in \eqref{sigma}, we get the RS1 model
\eqref{sigma_RS}, while putting $C = \pi \kappa r_c$, we come to the
RS-like scenario with the small curvature of the space-time (RSSC
model \cite{Giudice:2005}-\cite{Kisselev:2006}). What are main
features of the RSSC model in comparison with those of the RS1
model?

The interactions of the Kaluza-Klein (KK) gravitons
$h_{\mu\nu}^{(n)}$ with the SM fields on the TeV brane are given by
the effective Lagrangian density

\begin{equation}\label{Lagrangian}
\mathcal{L}_{\mathrm{int}} = - \frac{1}{\bar{M}_{\mathrm{Pl}}} \,
h_{\mu\nu}^{(0)}(x) \, T_{\alpha\beta}(x) \, \eta^{\mu\alpha}
\eta^{\nu\beta} - \frac{1}{\Lambda_\pi} \sum_{n=1}^{\infty}
h_{\mu\nu}^{(n)}(x) \, T_{\alpha\beta}(x) \, \eta^{\mu\alpha}
\eta^{\nu\beta} \;,
\end{equation}

\noindent were $\bar{M}_{\mathrm{Pl}} = M_{\mathrm{Pl}}/\sqrt{8\pi}$ is the
reduced Planck mass, $T^{\mu \nu}(x)$ is the energy-momentum tensor
of the SM fields. The coupling constant is equal to
\begin{equation}\label{Lambda_pi}
\Lambda_\pi = \bar{M}_5 \sqrt{\frac{\bar{M}_5}{\kappa}} \;.
\end{equation}
The hierarchy relation looks like
\begin{equation}\label{hierarchy}
\bar{M}_{\mathrm{Pl}}^2 = \frac{\bar{M}_5}{\kappa} \left[ e^{2\pi
\kappa r_c} - 1 \right] \Big|_{\kappa \pi r_c \gg 1} =
\frac{\bar{M}_5}{\kappa} \,e^{2\pi \kappa r_c} \;.
\end{equation}

The masses of the KK gravitons are proportional to the curvature
parameter $\kappa$ \cite{Kisselev:2005}

\begin{equation}\label{graviton_masses}
m_n = x_n \kappa \;, \quad n=1,2, \ldots \;,
\end{equation}

\noindent where $x_n$ are zeros of the Bessel function $J_1(x)$. If we put
$\kappa \ll \bar{M}_5 \sim 1$ TeV, the mass splitting will be small,
$\Delta m \simeq \pi \kappa$, and we come to an almost continuous
mass spectrum, similar to the mass spectrum of the ADD model
\cite{Arkani-Hamed:1998}. This is in contrast to the RS1 model, in
which the gravitons are heavy resonances with masses above one-few
TeV.

It is worth to stress that the RSSC model first proposed in
\cite{Giudice:2005} and after that developed in
refs.~\cite{Kisselev:2005}, \cite{Kisselev:2006},
\cite{Kisselev:2016} differs significantly from the 5-dimensional
space-time scenario with one flat extra dimension (ED) even for
small value of $\kappa$. To demonstrate this, consider the hierarchy
relation for the ADD model with $n$ EDs of the size $r_c$
\cite{Arkani-Hamed:1998}-\cite{Hamed2:1998}
\begin{equation}\label{hierarchy_ADD}
\bar{M}_{\mathrm{Pl}}^2 = (2\pi r_c)^n M_D^{2+n} \;,
\end{equation}
where $M_D$ is a fundamental gravity scale in $D=4+n$ dimensions.
For $n=1$ this equation is a particular case of the hierarchy
relation \eqref{hierarchy} in the limit $2\kappa \pi r_c \ll 1$.
However, the condition $2\kappa \pi r_c \ll 1$ means that
\cite{Kisselev:2006}
\begin{equation}\label{M5_kappa_ratio}
\frac{\bar{M}_5}{\kappa} \gg \left(
\frac{\bar{M}_{\mathrm{Pl}}}{\bar{M}_5} \right)^{\!2} .
\end{equation}
Inequality \eqref{M5_kappa_ratio} is satisfied if only $\kappa \ll
10^{-22}$ eV for $\bar{M}_5 = 1$ TeV. Thus, we conclude that the
RSSC model cannot be considered as a simple ``distortion'' of the
ADD model with one ED. The smallness of $\kappa$ means that $\kappa
\ll \bar{M}_5$, that is, $\kappa$ is very small in comparison with
the curvature in the original RS model \cite{Randall:1999} in which
$\kappa \sim M_{\mathrm{Pl}}$.

Let us calculate the scattering amplitude for the subprocess $\gamma
\gamma \rightarrow l^- l^+$ by adding $s$-channel KK graviton
exchange to the SM electromagnetic contribution, which is defined as

\begin{eqnarray}\label{KK_amplitude}
M_{KK} = \frac{1}{2 \Lambda_\pi^2} \sum_{n}
\,[\bar{u}(p_{1})\Gamma_{2}^{\mu\nu} v(p_{2}) \,
\frac{B_{\mu\nu\alpha\beta}}{\hat{s}-m^{2}_{n}} \,
\Gamma_{1}^{\alpha\beta\rho\sigma} e_{\rho}(k_{1})e_{\sigma}(k_{2})]
\;,
\end{eqnarray}

\noindent where $k_{1}, k_{2}$, $p_{1}, p_{2}$ and $e_{\rho}(k_{i})$ are
incoming photon, outgoing lepton momenta and polarization vectors of
photons. The coherent sum is over KK modes. Vertex functions
$\Gamma_{1}^{\alpha\beta\rho\sigma}$ for $KK$-- $\gamma\gamma$ and
$\Gamma_{2}^{\mu\nu}$ for $KK$-- $\ell\ell$ are given below

\begin{align}
\Gamma_{1}^{\alpha\beta\rho\sigma} &= -\frac{i}{2} \,[(k_{1}\cdot
k_{2})C^{\alpha\beta\rho\sigma}+
D^{\alpha\beta\rho\sigma}] \;, \label{Gamma_1} \\
\Gamma_{2}^{\mu\nu} &=-\frac{i}{8} \,
[\gamma^{\mu}(p^{\nu}_{1}-p^{\nu}_{2})+
\gamma^{\nu}(p^{\mu}_{1}-p^{\mu}_{2})] \;. \label{Gamma_2}
\end{align}

\noindent Explicit forms of the tensors $C^{\alpha\beta\rho\sigma}$ and
$D^{\alpha\beta\rho\sigma}$ are given by
eqs.~\eqref{tensor_C}-\eqref{tensor_D} in Appendix~A, while
$B_{\mu\nu\alpha\beta}$ is a tensor part of the graviton propagator
\begin{eqnarray}\label{graviton_propagator}
B_{\mu\nu\alpha\beta}=\eta_{\mu\alpha}\eta_{\nu\beta}+
\eta_{\mu\beta}\eta_{\nu\alpha}-\frac{2}{3} \,
\eta_{\mu\nu}\eta_{\alpha\beta} \;.
\end{eqnarray}

\noindent The total amplitude squared consists of electromagnetic, KK and
interference parts \cite{Atag:2009}

\begin{equation}\label{M_tot}
|M|^{2} = |M_{\mathrm{em}}|^{2} + |M_{\mathrm{\mathrm{KK}}}|^{2} +
|M_{\mathrm{int}}|^{2}  \;,
\end{equation}

\noindent where

\begin{align}
|M_{\mathrm{em}}|^{2} &= -2g^{4}_{e} \left[ \frac{\hat{s} +
\hat{t}}{\hat{t}} +
\frac{\hat{t}}{\hat{s} + \hat{t}} \right] , \label{M_em} \\
|M_{\mathrm{KK}}|^{2} &= \frac{1}{4} |S(\hat{s})|^{2}
\left[-\frac{\hat{t}}{8}
(\hat{s}^{3}+2\hat{t}^{3}+3\hat{t}\hat{s}^{2} +
4\hat{t}^{2}\hat{s}) \right] , \label{M_KK} \\
|M_{\mathrm{int}}|^{2} &= -\frac{1}{4} g^{2}_{e}
\,\mathrm{Re}S(\hat{s}) [ \hat{s}^{2}+2\hat{t}^{2} +
2\hat{s}\,\hat{t} ] \;. \label{M_int}
\end{align}

\noindent Here $\hat{s}$, $\hat{t}$ are Mandelstam variables of the subprocess
$\gamma \gamma \rightarrow l^- l^+$, and
$g^{2}_{e}=4\pi\alpha_{\mathrm{em}}$.

The $s$-channel contribution of the KK gravitons in \eqref{M_KK} and
\eqref{M_int} is given by the expression

\begin{equation}\label{S_def}
\mathcal{S}(s) =  \frac{1}{\Lambda_{\pi}^2} \sum_{n=1}^{\infty}
\frac{1}{s - m_n^2 + i \, m_n \Gamma_n} \;,
\end{equation}

\noindent were $\Gamma_n$ denotes the total width of the KK graviton with the
mass $m_n$. The sum \eqref{S_def} has been calculated in
ref.~\cite{Kisselev:2006}

\begin{equation}\label{S}
\mathcal{S}(s) = - \frac{1}{4\bar{M}_5^3 \sqrt{s}} \; \frac{\sin
(2A) + i \sinh (2\varepsilon)}{\cos^2 \!A + \sinh^2 \! \varepsilon }
\;,
\end{equation}

\noindent where

\begin{equation}\label{A_epsilon}
A = \frac{\sqrt{s}}{\kappa} \;, \qquad \varepsilon  = 0.045 \left(
\frac{\sqrt{s}}{\bar{M}_5} \right)^{\!\!3} .
\end{equation}

The virtual graviton exchange should lead to deviations from the SM
predictions both in a magnitude of the cross sections and in an
angular distribution of the final leptons because of the spin-2
nature of the gravitons.


\section{Numerical analysis} %

The KK graviton exchange studied in the previous Section should lead
to deviations from the SM in magnitudes both of differential cross
sections and of total cross sections for the photon-induced process
$pp \rightarrow p \mu^+ \mu^- p$ at the LHC which goes via
subprocess $\gamma\gamma \rightarrow \mu^+ \mu^-$. Our goal is to
calculate the dependence of these deviations on the parameters of
the RSSC model.

As it was mentioned in Introduction, the process can be detected by
using forward detectors CN-PPS (CMS-TOTEM Collaboration) and AFP
(ATLAS Collaboration). In what follows, we will consider two
acceptance regions of the forward detectors, $0.0015 < \xi < 0.5$
and $0.1 < \xi < 0.5$.

Before describing our numerical results, it is worth to make a few
remarks about using the cut imposed on the muon rapidities during
our calculations. In the c.m.s. of the colliding protons, two-muon
system $X$ moves with the rapidity

\begin{equation}\label{eta_X}
\eta_X = \frac{1}{2} \ln \frac{E_1}{E_2} = \frac{1}{2} \ln
\frac{\xi_1}{\xi_2} \;,
\end{equation}

\noindent where $E_{1,2} = \xi_{1,2} E$ are the energies of the photons,
$4E_1E_2 = \hat{s} = W^2$, $2E =\sqrt{s}$ is the invariant energy of
the $pp$ collision. The rapidities of the muons in the c.m.s. of two
\emph{photons} (= c.m.s of the system $X$) are equal to
$\eta_{\gamma\gamma}$ and $-\eta_{\gamma\gamma}$, respectively.
$\eta_{\gamma\gamma}$ depends on the scattering angle
$\theta_{\gamma\gamma}$ in the c.m.s. of the subprocess
$\gamma\gamma \rightarrow \mu^- \mu^+$

\begin{equation}\label{muon_rapididities_gg}
\eta_{\gamma\gamma} = \ln \left( \cot
\!\frac{\theta_{\gamma\gamma}}{2} \!\right) = \ln \!\frac{W +
\sqrt{W^2 - 4p_t^2}}{2p_t} \;.
\end{equation}

\noindent Correspondingly, the muon rapidities in the c.m.s. of the colliding
\emph{protons} are equal to

\begin{align}\label{muon_rapidities_pp}
\eta_{pp}^{(1)} &= \eta_{\gamma\gamma} + \eta_X \;, \nonumber
\\
\eta_{pp}^{(2)} &= -\eta_{\gamma\gamma} + \eta_X  \;.
\end{align}

\noindent It is clear that the rapidity cuts $|\eta_{pp}^{(1,2)}| <
\eta_{\max}$ are equivalent to the inequality
\begin{equation}\label{eta_gg_cut}
\eta_{\gamma\gamma} < \eta_{\max} - |\eta_X| \;.
\end{equation}
In our case $\eta_{\max} = 2.4$. Taking into account the cut imposed
on the transverse momenta of the muons, $|p_t| > p_{t,\min}$, one
can transform inequality \eqref{eta_gg_cut} into the bound on the
scattering angle of the muons in the c.m.s of two photons
\begin{equation}\label{cos_theta_ineq}
|\cos \theta_{\gamma\gamma}| < (\cos \theta)_{\max} \;,
\end{equation}
where
\begin{align}\label{cos_theta_bound}
(\cos \theta)_{\max} = \left\{
      \begin{array}{ll}
        0 \;, & \mathrm{if \ } |\eta_X| \geqslant \eta_{\max} \;, \\
        \min \displaystyle{ \!\left[ \sqrt{1 - \left(
        \frac{2p_{\mathrm{t}\min}}{W} \right)^{\!\!2}}, \tanh (\eta_{\max}) \right] } \;,
        & \mathrm{if \ } |\eta_X| < \eta_{\max} \;.
      \end{array}
    \right.
\end{align}

\noindent It is convenient to use the rapidity cut in such a form
\eqref{cos_theta_bound}. The reason is that the electromagnetic part
of the matrix element $|M|^2$ \eqref{M_em} depends only on $\cos
\theta_{\gamma\gamma}$, while other two terms in $|M|^2$,
\eqref{M_KK} and \eqref{M_int}, are functions of the variable $\cos
\theta_{\gamma\gamma}$ multiplied by functions of $\hat{s}$. Thus,
the corresponding integrals ($a =$ em, KK, int)
\begin{equation}\label{cs_sub}
\int\limits_{-(\cos \theta)_{\max}}^{(\cos \theta)_{\max}}
\!\!\frac{d\sigma_a}{d\cos \theta_{\gamma \gamma}} \,d \cos
\theta_{\gamma \gamma}
\end{equation}

\noindent can be calculated analytically.

The inequality $|\eta_X| < \eta_{\max}$ means that allowable values
of the muon transverse momenta are bounded from below
\begin{equation}\label{pt_lower_bound}
p_t > p_{t,0} = \frac{W_{\min}}{\cosh (\eta_{\max})} \;,
\end{equation}
where $W_{\min} = 2\xi_{\min} E$. Thus, for fixed $p_t > p_{t,0}$,
the following kinematical bounds take place
\begin{equation}\label{W_lower_bond}
\max (W_{\min}, 2p_t) \leqslant W \leqslant W_{\max} \;,
\end{equation}
where $W_{\max} = 2\xi_{\max} E$.

\begin{figure}[htb]
\begin{center}
\includegraphics[scale=0.60]{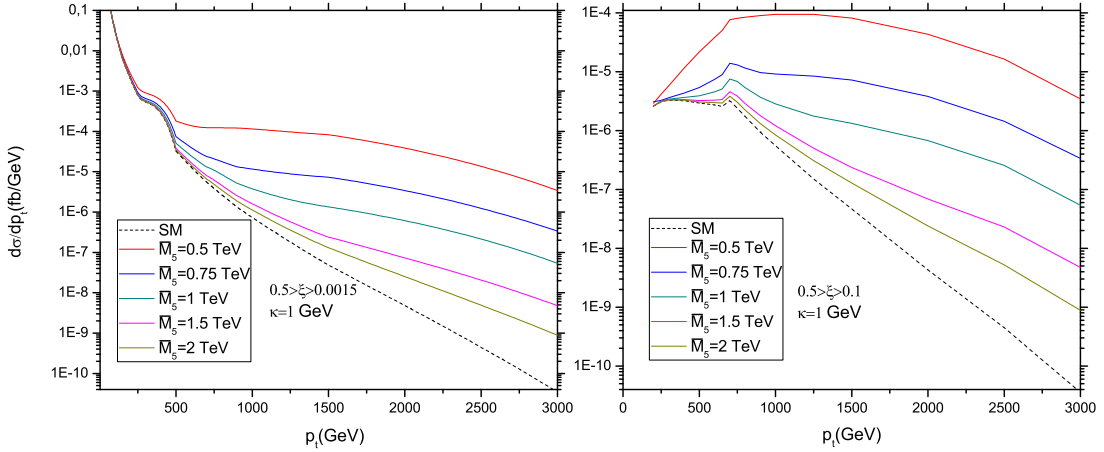}
\caption{The differential cross section for the process $pp
\rightarrow p\mu^+\mu^- p$ as a function of the transverse momenta
of the final muons for $\kappa = 1$ GeV and two acceptance regions.
The cut on the muon rapidities, $|\eta| < 2.4$, is imposed. Left
panel: $0.0015 < \xi < 0.5$. Right panel: $0.1 < \xi < 0.5$. Here
and below the dotted line denotes the SM contribution.}
\label{fig:ptd_M5}
\end{center}
\end{figure}

\begin{figure}[htb]
\begin{center}
\includegraphics[scale=0.60]{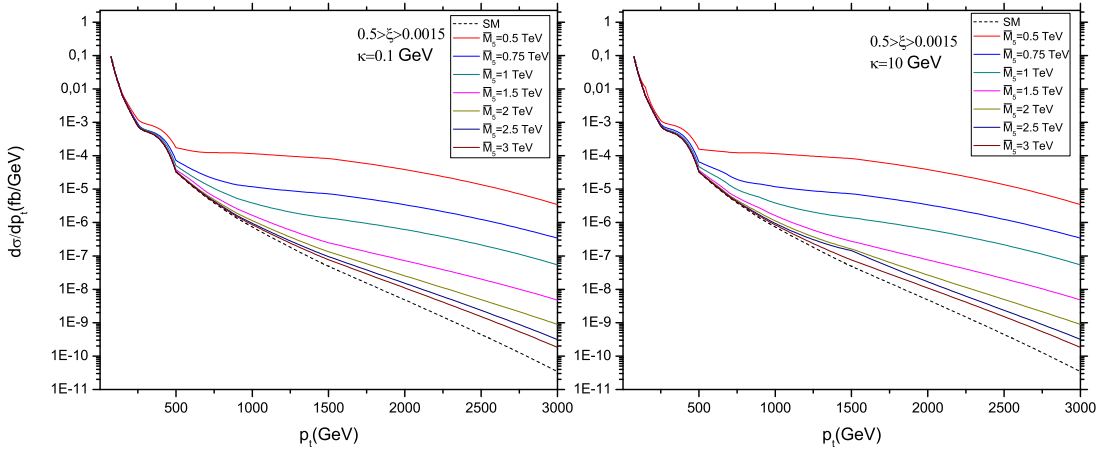}
\caption{The differential cross section for the process $pp
\rightarrow p\mu^+\mu^- p$ as a function of the transverse momenta
of the final muons for the acceptance region $0.0015 < \xi < 0.5$.
Left panel: $\kappa=0.1$ GeV. Right panel: $\kappa=10$ GeV.}
\label{fig:ptd_0015_kappa 01_10}
\end{center}
\end{figure}

\begin{figure}[htb]
\begin{center}
\includegraphics[scale=0.5]{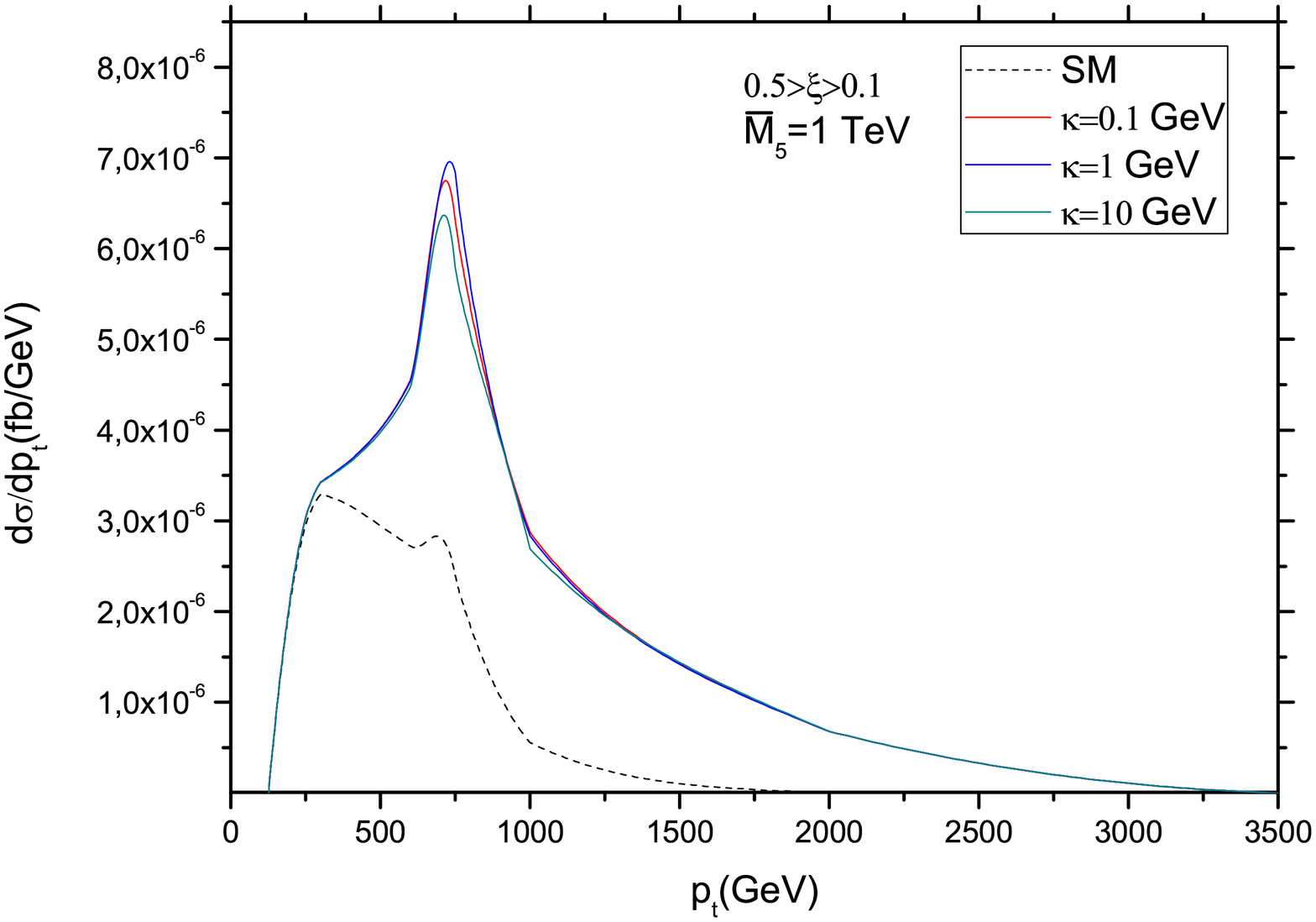}
\caption{The differential cross section for the process $pp
\rightarrow p\mu^+\mu^- p$ as a function of the transverse momenta
of the final muons for the acceptance region $0.1 < \xi < 0.5$ for
different values of $\kappa$.}
\label{fig:ptd_01_kappa}
\end{center}
\end{figure}

The results of our numerical calculations of the differential cross
sections $d\sigma/dp_t$ as a function of the transverse momenta of
the muons are presented in figs.~\ref{fig:ptd_M5}-\ref{fig:ptd_01_kappa}.
As one can see, $d\sigma/dp_t$
exceeds the SM contribution $d\sigma_{\mathrm{SM}}/dp_t$ for $p_t >
500$ GeV, and the difference between $d\sigma/dp_t$ and
$d\sigma_{\mathrm{SM}}/dp_t$ increases as $p_t$ grows. The effect is
more pronounced for small values of $\bar{M}_5$, for which
$d\sigma/dp_t$ becomes dominant already for $p_t \gtrsim 500\div600$
GeV.

Some comments should be made on a possible $\bar{M}_5$-dependence of
$d\sigma/dp_t$. One may naively expect from eq.~\eqref{S} that the
KK contribution to the differential cross section should be
\begin{equation}\label{KK_cs_expected}
\frac{d\sigma_{\mathrm{KK}}}{dp_t}\Big|_{\mathrm{naive}} \sim
\frac{1}{\bar{M}_5^6} \;.
\end{equation}
Nevertheless, the results of our numerical calculations shows that
\begin{equation}\label{KK_cs_num}
\frac{d\sigma_{\mathrm{KK}}}{dp_t} \sim \frac{1}{\bar{M}_5^3} \;,
\end{equation}
at small and moderate values of $p_t$. As a result, the differential
cross section $d\sigma/dp_t$ follows this $\bar{M}_5$-dependence. At
large $p_t$ the $\bar{M}_5$-dependence tends to the form
$\bar{M}_5^{-6}$. It can be explained as follows
\cite{Kisselev:2008}-\cite{Kisselev:2013}. In the RSSC model the
invariant part of the scattering amplitude \eqref{S} is a sum of
rather sharp resonances whose widths are proportional to
$\kappa^4/\bar{M}_5^3$. The contribution of one resonance can be
estimated as \cite{Kisselev:2008}-\cite{Kisselev:2013}
\begin{equation}\label{KK_cs_obtained}
\frac{d\sigma_{\mathrm{KK}}}{dp_t}\Big|_{\mathrm{one \ res}} \sim
\frac{\kappa}{\bar{M}_5^3 W} \;.
\end{equation}
Taking into account that the total number of the graviton resonances
which contribute to the differential cross section $\sim W/\kappa$,
we come to eq.~\eqref{KK_cs_num}.

From eqs.~\eqref{completeprocess} with using
dimensionless parameter $z = \sqrt{W^2/4p_t^2 - 1}$ instead of
variable $W$, the SM differential cross section can be obtained
analytically as follows
\begin{align}\label{sigma_SM}
\frac{d\sigma_{\mathrm{SM}}}{dp_t} &= \frac{ge^4}{16\pi }
\frac{1}{p_t^3} \, \int\limits_{z_{\min}}^{z_{\max}} \!\!dz
\frac{2z^2 + 1}{(z^2+1)^{5/2}} \int \! \frac{d\xi_2}{\xi_2} \int
\!\frac{dQ_1^2}{Q_1^2} \int \!\frac{d
Q_2^2}{Q_2^2} \nonumber \\
&\times \tilde{f}_1 \!\!\left( \frac{W^2(z)}{4E^2\xi_2 }\right)
\!\tilde{f}_2(\xi_2, Q_2^2) \;,
\end{align}
where
\begin{equation}\label{photon_distr}
\tilde{f}_i(\xi, Q^2) =  \xi_i Q_i^2 \,\frac{dN_\gamma(\xi,
Q^2)}{\xi_i dQ_i^2}
\end{equation}
is a reduced dimensionless photon distribution($i=1,2$) and
\begin{equation}\label{z_min_max}
z_{\min} = \sqrt{\frac{W_{\min}^2}{4p_t^2} - 1} \;, \quad z_{\max} =
\sqrt{\frac{W_{\max}^2}{4p_t^2} - 1} \;,
\end{equation}
\begin{equation}\label{W_min_max}
W_{\min} = \max (\xi_{\min} 2E , 2p_t) \;, \quad W_{\max} =
\xi_{\max} 2E \;.
\end{equation}
The $p_t$-dependence of the integrand in \eqref{sigma_SM} results in
(after integrations in all variables) both the small dip (around the
$600$ GeV) and two maxima (first is around the $300$ GeV and second
is around the $700$ GeV) for the SM contribution as seen from
fig.~\ref{fig:ptd_01_kappa}.

Figure \ref{fig:ptd_01_kappa} demonstrates to us that the
differential cross section is almost independent of the curvature
parameter $\kappa$, with the exception of its weak dependence on
$\kappa$ around the point $p_t = 750$ GeV.  Thus, we can put limits
on the fundamental gravity scale $\bar{M}_5$ regardless of the
parameter $\kappa$ (provided $\kappa \ll \bar{M}_5$ is satisfied).
This is in contrast to the RS1 model \cite{Randall:1999} in which
all cross sections depend substantially on the ratio $\beta =
\kappa/M_{\mathrm{Pl}}$.

\begin{figure}[htb]
\begin{center}
\includegraphics[scale=0.5]{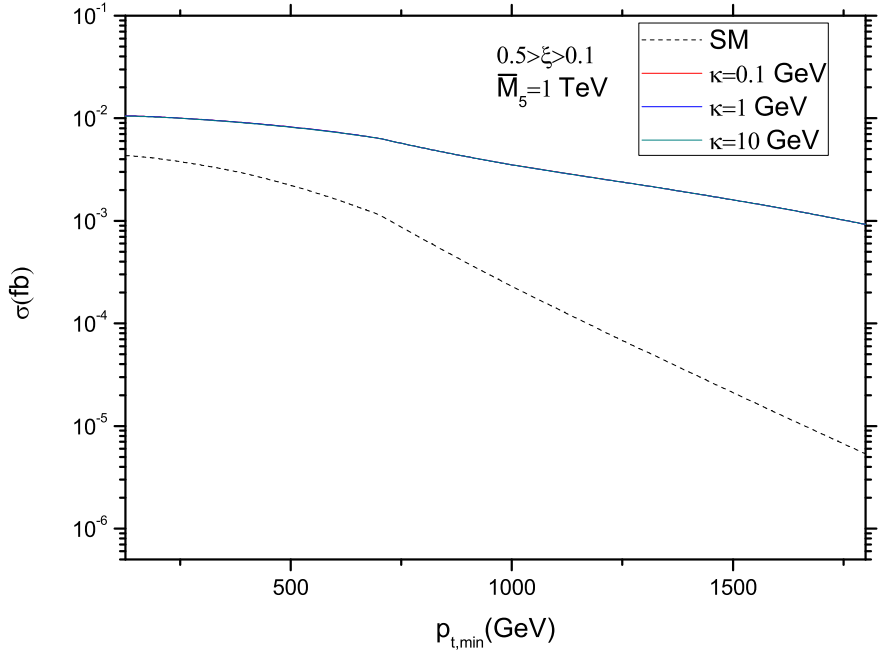}
\caption{The total cross section for the process $pp \rightarrow
p\mu^+\mu^- p$ as a function of the minimal transverse momenta of
the final muons $p_{t,\min}$  for the acceptance region $0.1 < \xi <
0.5$ for different values of $\kappa$.} \label{fig:cs_01_kappa}
\end{center}
\end{figure}

\begin{figure}[htb]
\begin{center}
\includegraphics[scale=0.60]{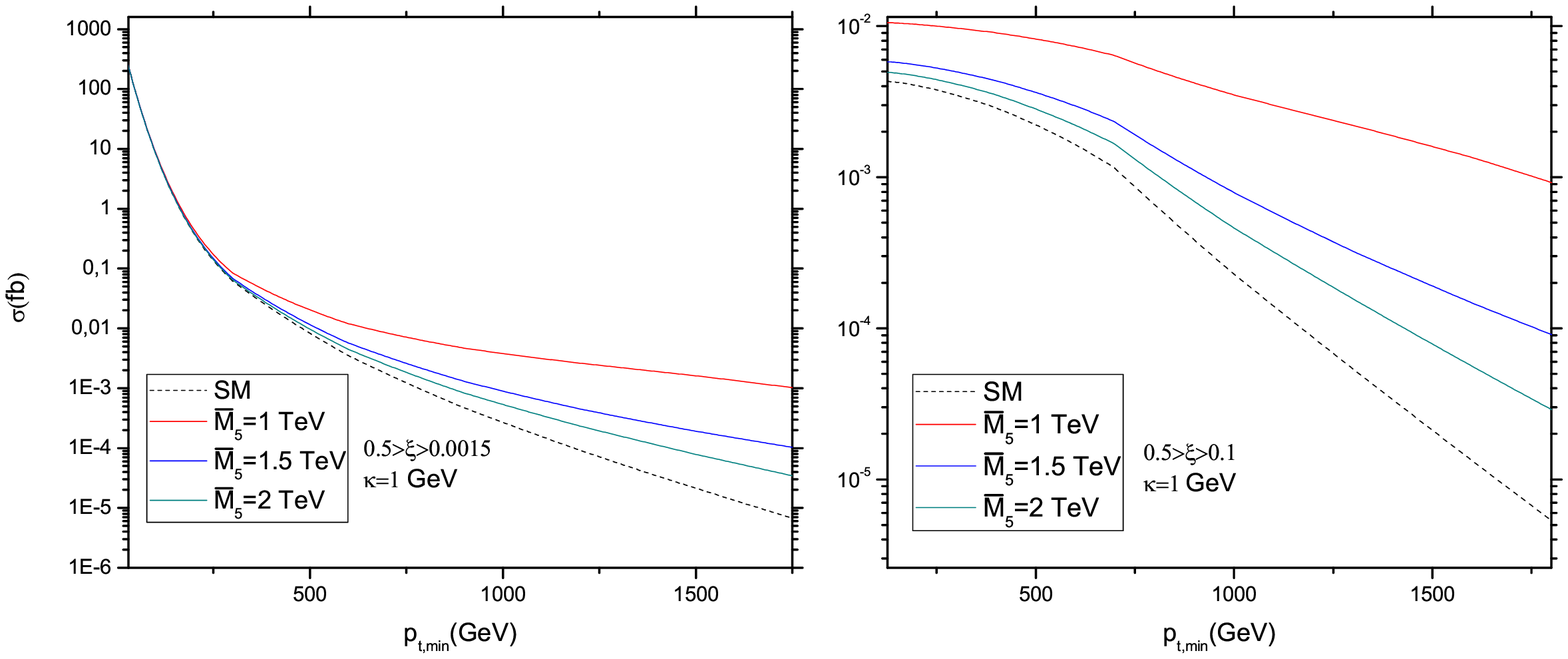}
\caption{The total cross section for the process $pp \rightarrow
p\mu^+\mu^- p$ as a function of the minimal transverse momenta of
the final muons $p_{t,\min}$  for different values of $\bar{M}_5$.
Left panel: $0.0015 < \xi < 0.5$. Right panel: $0.1 < \xi < 0.5$.}
\label{fig:cs_01_M5}
\end{center}
\end{figure}

The next two figures \ref{fig:cs_01_kappa}-\ref{fig:cs_01_M5} shows
us the total cross sections with and without KK graviton exchange
versus the minimal transverse momentum of the final muons
$p_{t,\min}$. The comparison with the pure SM predictions is also
given. From all said above it is not surprising that the quantity
$\sigma(p_t > p_{t,\min})$ does not depend on $\kappa$ (see
fig.~\ref{fig:cs_01_kappa}). For both acceptance regions, its
deviation from the SM gets higher as $p_{t,\min}$ grows (see
fig.~\ref{fig:cs_01_M5}). When the two figures are compared, we can
see that the $0.1<\xi<0.5$ case has almost the same behaviour as the
case $0.0015<\xi<0.5$ with $p_{t,\min}\sim 500$ GeV. Moreover, for
the $0.1<\xi<0.5$ acceptance region, as the $p_{t,\min}$ changes,
the cross sections almost do not change if $0<p_{t,\min}<500$.
Therefore, it can be said that a high value of $\xi_{\min}$ mimics a
high value of $p_{t,\min}$.

\begin{figure}[htb]
\begin{center}
\includegraphics[scale=0.60]{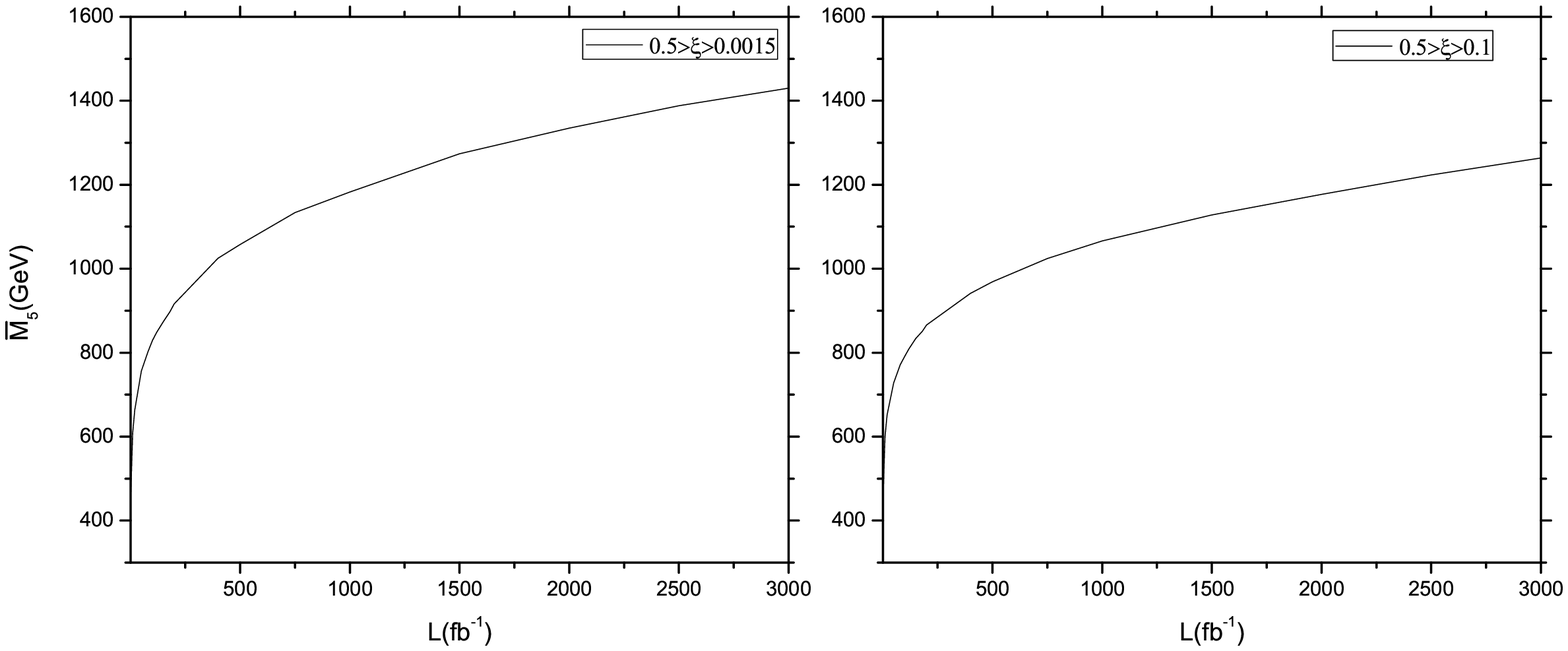}
\caption{The 95 C.L.\% search limits for the reduced 5-dimensional
gravity scale $\bar{M}_5$ as a function of the integrated LHC
luminosity for $0.0015>\xi>0.5$ with $p_t>500$ GeV and $0.1<\xi<0.5$
with $p_t>30$ GeV. The rapidity cut of 2.4 on the muon rapidities
are imposed.} \label{fig:SS}
\end{center}
\end{figure}

In this motivation, we have obtained the limits on the $\bar{M}_5$
for two cases: $0.0015<\xi<0.5$ for $p_t>500$ GeV and $0.1<\xi<0.5$
for $p_t>30$ GeV. In sensitivity analysis, the likelihood method are
used. In this method, it is assumed that the entire number of
background events in every signal location respects the normal
distribution with a fractional uncertainty. The statistical
significance is obtained as follows \cite{SS},

\begin{eqnarray}
SS=\sqrt{2[(S+B)\ln(1+S/B)-S]} \;,
\label{Eq:SS}
\end{eqnarray}

\noindent where $S$, $B$ are the signal and background events
number, respectively. Here, we have assumed that the uncertainty of
the background is negligible. Using our results on the cross
sections, we have calculated 95\% C.L. bounds on $\bar{M}_5$ for two
acceptance regions, $0.0015 < \xi < 0.5$ with $p_t>500$ GeV and $0.1
< \xi < 0.5$ with $p_t>30$ GeV with using eq.(~\ref{Eq:SS}). The
bounds are presented in fig.~\ref{fig:SS} as a function of the
integrated LHC luminosity. Here, we assumed that the $\kappa=1$ GeV.
Note that the cut $|\eta| < 2.4$ was imposed on the rapidities on
the final muons. The results are presented in fig.~\ref{fig:SS}
which have nearly the same behavior for the two acceptances regions.
It can be see that 95\% C.L. sensitivity of $\bar{M}_5$ is about
$1430$ GeV for the $0.0015 < \xi < 0.5$ and $1265$ GeV for the $0.1
< \xi < 0.5$ when the LHC integrated luminosity values are equal to
$3000$ fb$^{-1}$.

Our bounds on the 5-dimensional gravity scale $\bar{M}_5$ are rather
low in comparison with the LHC bounds on $D$-dimensional scale in
the ADD model $M_D$. In this regard, we would like to emphasize the
following. The LHC bounds on $M_D$ cannot be directly applied to the
gravity scale of the RSSC model. As was shown above (see comments
after eq.~\eqref{graviton_masses}), this model cannot be regarded as
a small distortion of the ADD model. Moreover, in the ADD model the
number of EDs should be $n \geqslant 2$, while in the RSSC scenario
we deal with one extra dimension, $n=1$. As for the original RS
model, it has $\bar{M}_5 \sim \kappa \sim M_{\mathrm{Pl}}$, and
bounds are put on the ratio $\kappa/\bar{M}_5$ and mass of the
lightest KK graviton $m_1$. We consider the diphoton production in
the photon induced process at the LHC as another means of seeing
effects of low $M_5$ in the RS-like scenario with the small
curvature.

A framework that allows a self-consistent description of quantum
gravity is string theory \cite{Polchinski:98}. In the presence of
extra dimensions, the string scale $M_s$ and the fundamental Planck
scale $M_P$ could be as low as $\sim$ TeV
\cite{Arkani-Hamed:1998}-\cite{Hamed2:1998}. The TeV-gravity theory
has four new types of particles (higher-dimensional gravitons,
low-lying string excitations, string balls (SB) \cite{Dimopoulos:02}
and black holes (BH) \cite{Banks:99}-\cite{Giggings:02}) and three
associated mass scales ($M_P$, $M_s$, and $M_s/g_s^2$, where $g_s$
is the string coupling).

If $g_s \ll 1$ (perturbative string theory), there is a separation
between these scales. In the type I string theory with D-branes  the
scales $M_P$ and $M_s$ are related with each other
\cite{Hamed2:1998}. Taking into account that in some respect (small
curvature of the space-time, almost continuous spectrum of massive
gravitons) the RSSC model is similar to the ADD model with one extra
dimension, we have
\begin{equation}\label{MD_Ms_relation}
\frac{M_s}{M_P} = (4\pi \alpha)^{2/3} \;,
\end{equation}
where $\alpha = g^2/(4\pi)$ is the gauge coupling, and the
proportionality constant follows from the convention of
ref.~\cite{Giudice:02}. For instance, we find that $M_s/M_P$ is
equal to $0.25(1.16)$ for $\alpha = 0.01(0.1)$. Typically,
\begin{align}\label{scales}
M_s < M_P < \frac{M_s}{g_s^2} \;, \nonumber \\
M_s \ll M_{\mathrm{SB}} \ll \frac{M_s}{g_s^2} \;, \nonumber \\
\frac{M_s}{g_s^2} \ll M_{\mathrm{BH}} \;,
\end{align}
where $M_{\mathrm{SB}}$ is the string ball mass, while
$M_{\mathrm{BH}}$ is the mass of the black hole .

If $g_s \sim 1$, the mass scales $M_s$ and $M_s/g_s^2$ coincide and
calculability of the string theory is lost. The black holes are
expected to dominate the dynamics above $M_s$.

However, neither black holes with masses less than 10.1 TeV nor
string balls with masses less than 9.5 TeV were seen at the LHC
\cite{CMS:BH_min}. That is why, we assumed an absence of
string-gravity corrections in our calculations.


\section{Conclusions} %

Photon-induced exclusive processes $pp \rightarrow p \gamma\gamma p
\rightarrow p X p$ are of great importance for high-energy physics.
They provide one with unique precision measurements of the
electroweak sector of the SM. They also allow us to study physics
beyond the SM. For instance, (semi)exclusive $WW$ production by
photon-photon interactions is very sensitive to quartic gauge
anomalous couplings.

The dilepton production at the electroweak scale have been studied
both at the Tevatron and LHC colliders. However, all previous
experiments was done without a proton tag. Recently, the dimuon
production in the process $pp \rightarrow p \gamma\gamma p^\ast
\rightarrow p \mu^+ \mu^- p^\ast$ have been studied with the
CMS-TOTEM forward detector CT-PPS using measurements based on the
integrated luminosity of 10 fb$^{-1}$ at 13 TeV
\cite{CMS-TOTEM:2018}. 12 events with $m_{\mu\mu} > 110$ GeV
matching forward detector kinematics were observed. This result is
the first observation of proton-tagging $\gamma\gamma$ electroweak
collisions.

In ref.~\cite{Atag:2009} a potential of the photon-induced dilepton
final states at the LHC for a phenomenology of two models with the
extra dimensions was investigated. The constraints both on the
fundamental gravity scale $M_D$ in the ADD model and on the pair
$\beta $ -- $m_1$ in the RS1 model (where $\beta =
\kappa/\bar{M}_5$, $m_1$ is a mass of the lightest graviton), were
derived \cite{Atag:2009}.

In the present paper we have studied the photon-induced production
of the muon pair at the LHC for 14 TeV in the RSSC model with the
warped extra dimension and small curvature
\cite{Giudice:2005}-\cite{Kisselev:2006}. For two acceptance
regions, $0.0015<\xi<0.5$ and $0.1<\xi<0.5$, where $\xi$ is the
proton energy fraction loss, the distributions in the muon
transverse momenta $p_t$ are calculated as a function of the reduced
fundamental gravity scale $\bar{M}_5$ and curvature parameter
$\kappa$. It is shown that the deviation from the SM gets higher as
$p_t$ grows. The obtained cross sections $\sigma(p_t > p_{t,\min})$
almost do not change in the region $0 < p_{t,\min} < 500$ GeV for
the case $0.1<\xi<0.5$. It means that the high value of $\xi_{\min}$
mimics the high value of $p_{t,\min}$. The 95\% L.C. discovery
limits on $\bar{M}_5$ are obtained for the acceptance region
$0.0015<\xi<0.5$ with $p_t>500$ GeV and for $0.1<\xi<0.5$ with
$p_t>30$ GeV. Note that the cut $|\eta|< 2.4$ was imposed on the
rapidities of the final muons.

Let us stress that our limits on $\bar{M}_5$ do not depend on the
curvature parameter $\kappa$. In the RSSC model this fact also takes
place for other processes, provided the inequality $\kappa \ll
\bar{M}_5$ is satisfied \cite{Kisselev:2008}. Such a weak dependence
on $\kappa$ for $\kappa \ll \bar{M}_5$ can be understood as follows.
Consider a contribution of a gravity resonance with the mass $m_0 =
\sqrt{s \tau_0}$ to the sum $\mathcal{S}(s)$ \eqref{S_def}. Its real
part has two peaks with opposite signs which cancel each other. As
for the imaginary part of this resonance, its height is proportional
to $1/\varepsilon_0$, where
\begin{equation}\label{epsilon_0}
\varepsilon_0 = \frac{\eta}{2} \left( \frac{\sqrt{s
\tau_0}}{\bar{M}_5} \right)^{\!3} ,
\end{equation}
while its width is equal to $2\delta_0$, where
\begin{equation}\label{delta_0}
\delta_0 = \eta \frac{\kappa s \tau_0}{\bar{M}_5^3} \;.
\end{equation}
The total number of the graviton resonances which contribute to the
differential cross section $d\sigma/dp_t^2$ is proportional to
$N/\kappa$. As a result, we find that the gravity contribution to
the differential cross section is proportional to
\begin{equation}\label{KK_to_dif_cs}
s^2 \!\left[ \frac{1}{\bar{M}_5^3 \sqrt{s} \,\varepsilon_0}
\right]^{\!2} \!\delta_0 N \sim \frac{1}{\bar{M}_5^3 \sqrt{s}} \;.
\end{equation}
Thus, the smallness of the coupling constant $1/\Lambda_\pi^2 =
\kappa/\bar{M}_5^3$ in (24) is compensated by a large number of
gravitons $N = \sqrt{s}/\kappa$ which give significant contribution
to $\mathcal{S}(s)$.

It is in contrast to the RS1 model, in which $\kappa \sim \bar{M}_5
\sim M_{\mathrm{Pl}}$. As a result, the RS1 discovery limits on
$m_1$ for all processes, including photon-induced collisions at the
LHC \cite{Atag:2009}, depend on a chosen value of $\kappa$.



\setcounter{equation}{0}
\renewcommand{\theequation}{A.\arabic{equation}}

\section*{Appendix A}

Symbols $C^{\alpha\beta\rho\sigma}$ and $D^{\alpha\beta\rho\sigma}$
in eq.~\eqref{Gamma_1} are defined as follows
\begin{align}
C^{\alpha\beta\rho\sigma} &= \eta^{\alpha\rho}\eta^{\beta\sigma}
+\eta^{\alpha\sigma}\eta^{\beta\rho} -
\eta^{\alpha\beta}\eta^{\rho\sigma} \;, \label{tensor_C} \\
D^{\alpha\beta\rho\sigma} &=\eta^{\alpha\beta}
k^{\sigma}_{1}k^{\rho}_{2} - (\eta^{\alpha\sigma}
k^{\beta}_{1}k^{\rho}_{2} + \eta^{\alpha\rho}
k^{\sigma}_{1}k^{\beta}_{2} - \eta^{\rho\sigma}
k^{\alpha}_{1}k^{\beta}_{2}) \nonumber \\
&- (\eta^{\beta\sigma} k^{\alpha}_{1}k^{\rho}_{2} + \eta^{\beta\rho}
k^{\sigma}_{1}k^{\alpha}_{2} - \eta^{\rho\sigma}
k^{\beta}_{1}k^{\alpha}_{2}) \;. \label{tensor_D}
\end{align}





\begin{thebibliography}{99}

\bibitem{afp} L. Adamczyk, et al., Tech. Rep. CERN-LHCC-2015-009.
ATLAS-TDR-024, May, 2015.

\bibitem{totem} M. Albrow et al., CMS-TOTEM, Tech. Rep. CERN-LHCC-2014-021. TOTEM-TDR-003.
CMS-TDR-13, September, 2014.

\bibitem{afp1} ATLAS Collaboration,  Tech. Rep. CERN-LHCC-2011-
012. LHCC-I-020, (2011).
\bibitem{afp2} L. Adamczyk {\it et al.}, Tech. Rep. ATLCOM-
LUM-2011-006, CERN, 2011.

\bibitem{ttm1} G. Antchev {\it et al.} (TOTEM Collaboration),  EPL 96 21002, (2011).
\bibitem{ttm2} G. Antchev {\it et al.} (TOTEM Collaboration),  EPL 98 31002, (2012).
\bibitem{ttm3} G. Antchev {\it et al.} (TOTEM Collaboration), EPL \textbf{101}, 21002 (2013).

\bibitem{albrow} M. Albrow {\it et al.}, (FP420 R and D Collaboration), JINST 4, T10001 (2009).
\bibitem{albrow1} M.G. Albrowa, T.D. Coughlin and J.R. Forshaw, Prog. Part. Nucl. Phys. 65, 149 (2010).

\bibitem{cdf1} T. Aaltonen {\it et al.}, (CDF Collaboration), Phys. Rev. Lett. {\bf 102}, 242001
(2009).

\bibitem{cdf2} T. Aaltonen {\it et al.}, (CDF Collaboration), Phys. Rev. Lett. {\bf 102}, 222002
(2009).
\bibitem{ch1} S. Chatrchyan {\it et al.}, (CMS Collaboration), JHEP {\bf1201}, 052 (2012).
\bibitem{ch2} S. Chatrchyan {\it et al.}, (CMS Collaboration), JHEP {\bf1211}, 080 (2012).

\bibitem{ath1} G. Aad {\it et al.}, (ATLAS Collaboration)Phys. Lett. B {\bf749}, 242, (2015)
\bibitem{ath2} G. Aad {\it et al.}, (ATLAS Collaboration), JHEP  {\bf08} 009
(2016).

\bibitem{lhc2} K. Piotrzkowski, Phys. Rev. D {\bf 63}, 071502(R) (2001).
\bibitem{lhc4} V. Goncalves and M. Machado, Phys. Rev. D {\bf
75}, 031502(R) (2007).
\bibitem {inanc} \.{I}. \c{S}ahin and S.C. \.{I}nan, JHEP {\bf09}, 069 (2009).
\bibitem {inan} S.C. \.{I}nan, Phys. Rev. D {\bf 81}, 115002 (2010).
\bibitem {bil} S. Ata\u{g} and A. Billur, JHEP {\bf 11} 060 (2010).
\bibitem {bil2} \.{I}. \c{S}ahin and A. A. Billur, Phys. Rev. D {\bf 83}, 035011 (2011).
\bibitem {kok} \.{I}. \c{S}ahin and M. K\"{o}ksal, JHEP {\bf 11}, 100 (2011).
\bibitem {inan2} S.C. \.{I}nan and A. A. Billur, Phys. Rev. D {\bf 84}, 095002 (2011).
\bibitem {gru} R. S. Gupta, Phys. Rev. D {\bf 85}, 014006 (2012).
\bibitem {inanc2} \.{I}. \c{S}ahin, Phys. Rev. D {\bf 85}, 033002 (2012).
\bibitem{ban} B. \c{S}ahin and A. A. Billur,  Phys.Rev. D {\bf 85} 074026 (2012).
\bibitem {epl} L.N. Epele {\it et al.}, Eur. Phys. J. Plus {\bf127}, 60 (2012).
\bibitem {inanc3} \.{I}. \c{S}ahin and B. \c{S}ahin, Phys. Rev. D {\bf 86}, 115001 (2012).
\bibitem {bil4} A.A. Billur, Europhys. Lett. {\bf101}, 21001 (2013).
\bibitem {inanc4} \.{I}. \c{S}ahin {\it et al.}, Phys.Rev. D {\bf 88} 095016 (2013).
\bibitem{hao1} H. Sun, C. X. Yue, Eur. Phys. J. C 74, 2823 (2014).
\bibitem{hao2} H. Sun, Nucl. Phys. B 886, 691 (2014).
\bibitem{ins} \.{I}. \c{S}ahin {\it et al.}, Phys.Rev. D {\bf 91} 035017 (2015).
\bibitem{kok2} M. K\"{o}ksal and S.C. \.{I}nan, Adv. High Energy Phys. 2014, 315826 (2014)
\bibitem{fic1} S. Fichet, JHEP {\bf 1704}, 088 (2017).
\bibitem{fic2} C. Baldenegro {\it et al.}, JHEP {\bf 1706}, 142 (2017).

\bibitem{budnev} V. Budnev, I. Ginzburg, G. Meledin and V.
Serbo, Phys. Rep. {\bf 15}, 181 (1975).
\bibitem{baur} G. Baur {\it et al.}, Phys. Rep. {\bf 364}, 359 (2002).

\bibitem{Randall:1999}
L.~Randall and R.~Sundrum, Phys. Rev. Lett. \textbf{83}, 3370
(1999).
%
\bibitem{Arkani-Hamed:1998} N.~Arkani-Hamed, S.~Dimopoulos and G.~Dvali, Phys. Lett. B
\textbf{429}, 263 (1998);
\bibitem{Hamed:1998} N.~Arkani-Hamed, S.~Dimopoulos and G.~Dvali, Phys. Rev. D \textbf{59}, 086004 (1999);
\bibitem{Hamed2:1998} I.~Antoniadis, N.~Arkani-Hamed, S.~Dimopoulos and G.~Dvali, Phys.
Lett. B \textbf{436}, 257 (1998).
%
\bibitem{Kisselev:2016}
A.V.~Kisselev, Nucl. Phys. B \textbf{909}, 218 (2016).
%
\bibitem{Giudice:2005}
G.F.~Giudice, T.~Plehn and A.~Strumia, Nucl. Phys. B \textbf{706},
455 (2005).
%
\bibitem{Kisselev:2005}
A.V.~Kisselev and V.A.~Petrov, Phys. Rev. D \textbf{71}, 124032
(2005).
\bibitem{Kisselev:2006}
A.V.~Kisselev, Phys. Rev. \textbf{D 73}, 024007 (2006).
%
\bibitem{Atag:2009}
S.~Ata\v{g}, S.C.~\.{I}nan, \.{I}~\c{S}ahin, Phys. Rev. D
\textbf{80}, 075009 (2009).
\bibitem{Kisselev:2008}
A.V.~Kisselev, JHEP \textbf{0809}, 039 (2008).
%
\bibitem{Kisselev:2013}
A.V.~Kisselev, JHEP \textbf{1304}, 025 (2013).
\bibitem{SS}  G. Cowan, K. Cranmer, E. Gross, O. Vitells, Eur. Phys. J. C \textbf{71},
1554 (2011).
%
\bibitem{Polchinski:98}
J.~Polchinski, \emph{String theory}, Cambridge University Press,
1998, Vols.~I and II.
%
\bibitem{Dimopoulos:02}
S.~Dimopoulos and R.~Emparan, Phys. Lett. \textbf{B526}, 393 (2002).
%
\bibitem{Banks:99}
T.~Banks and W.~Fischler, hep-th/9906038.
\bibitem{Dimopoulos:01}
S.~Dimopoulos and G.~Landsberg, Phys. Rev. Lett. \textbf{87}, 161602
(2001).
\bibitem{Giggings:01}
S.B.~Giggings, eConf \textbf{C010630} (2001) P328.
\bibitem{Giggings:02}
S.B.~Giddings and S.~Tomas, Phys. Rev. D \textbf{65}, 056010 (2002).
%
\bibitem{Giudice:02}
G.F.~Giudice, R.~Rattazzi and J.D.~Wells, Nucl. Phys. B
\textbf{630}, 293 (2002).
%
\bibitem{CMS:BH_min}
A.M.~Sirunyan \emph{et al.} (CMS Collaboration), Phys. Lett. B
\textbf{774}, 279 (2017).
%
\bibitem{CMS-TOTEM:2018}
A.M.~Sirunyan \emph{et al.} (CMS-TOTEM Collaboration), JHEP
\textbf{07}, 153 (2018).
%
\end{thebibliography}
\end{document}